\documentstyle[twoside,fleqn,espcrc2,epsf]{article}

\def\bfone{\relax{\rm 1\kern-.35em 1}}
\def\bfzero{\relax{\rm I\kern-.18em 0}}
\def\inbar{\vrule height1.5ex width.4pt depth0pt}
\def\IC{\relax\,\hbox{$\inbar\kern-.3em{\rm C}$}}
\def\ID{\relax{\rm I\kern-.18em D}}
\def\IF{\relax{\rm I\kern-.18em F}}
\def\IH{\relax{\rm I\kern-.18em H}}
\def\II{\relax{\rm I\kern-.17em I}}
\def\IL{\relax{\rm I\kern-.17em L}}
\def\IN{\relax{\rm I\kern-.18em N}}
\def\IP{\relax{\rm I\kern-.18em P}}
\def\IQ{\relax\,\hbox{$\inbar\kern-.3em{\rm Q}$}}
\def\IR{\relax{\rm I\kern-.18em R}}
\def\IG{\relax\,\hbox{$\inbar\kern-.3em{\rm G}$}}
\font\cmss=cmss10 \font\cmsss=cmss10 at 7pt
\def\ZZ{\relax\ifmmode\mathchoice
{\hbox{\cmss Z\kern-.4em Z}}{\hbox{\cmss Z\kern-.4em Z}}
{\lower.9pt\hbox{\cmsss Z\kern-.4em Z}}
{\lower1.2pt\hbox{\cmsss Z\kern-.4em Z}}\else{\cmss Z\kern-.4em
Z}\fi}

\def\cA{{\cal A}}

 \def\cM{{\cal M}}
 
 \def\cQ{{\cal Q}}
 
\def\Coe#1.#2.{{#1\over #2}}

\def\coe#1.#2.{\relax{\textstyle {#1 \over #2}}\displaystyle}

\def\to{\rightarrow}
\def\notin{\hbox{{$\in$}\kern-.51em\hbox{/}}}

\def\IE{\relax{{\rm I\kern-.18em E}}}

\def\IGam{\relax{{\rm I}\kern-.18em \Gamma}}

\def\IA{\relax{\hbox{{\rm A}\kern-.82em {\rm A}}}}

\newcommand{\be}{\begin{equation}}
\newcommand{\ee}{\end{equation}}
\newcommand{\ba}{\begin{eqnarray}}
\newcommand{\ea}{\end{eqnarray}}
\newtheorem{definizione}{Definition}[section]
\newcommand{\bd}{\begin{definizione}}
\newcommand{\ed}{\end{definizione}}
\newtheorem{teorema}{Theorem}[section]
\newcommand{\bth}{\begin{teorema}}
\newcommand{\eth}{\end{teorema}}
\newtheorem{lemma}{Lemma}[section]
\newcommand{\blem}{\begin{lemma}}
\newcommand{\elem}{\end{lemma}}
\newcommand{\brr}{\begin{array}}
\newcommand{\err}{\end{array}}

\newtheorem{corollario}{Corollary}[section]
\newcommand{\bcorol}{\begin{corollario}}
\newcommand{\ecorol}{\end{corollario}}

\newcommand{\AmS}{{\protect\the\textfont2
  A\kern-.1667em\lower.5ex\hbox{M}\kern-.125emS}}
\title{Partial $N=2 \to N=1$ Local Supersymmetry Breaking\\
and Solvable Lie Algebras }
\author{Pietro
Fr\'e,\address{Dipartimento di Fisica Teorica, Universit\`a di 
Torino, \\ Via P. Giuria 1, I-10125 TORINO, Italy},
Luciano Girardello\address{Dipartimento di Fisica, Universit\'a
di Milano, via Celoria 6, I-20133 Milano,\\ and Istituto Nazionale di 
Fisica Nucleare (INFN)--Sezione di Milano, Italy}, Igor Pesando\address{Dipartimento di Fisica Teorica, Universit\`a di 
Torino, \\ Via P. Giuria 1, I-10125 TORINO, Italy},
Mario Trigiante\address{International School of Advanced Studies
 (ISAS), via Beirut 2--4, I-34100 Trieste,\\ and Istituto Nazionale di 
Fisica Nucleare (INFN)--Sezione di Trieste, Italy}%
\thanks{Supported in part by EEC under TMR contract ERBFMRX-CT96-0045 } } 
\begin{document}
\begin{abstract}
Generic partial supersymmetry breaking of $N=2$ supergravity
with zero vacuum energy and
with surviving unbroken arbitrary gauge groups
is exhibited. Specific examples are given.
\end{abstract}
\maketitle
\section{Introduction}
\label{intro}
A great number of considerable results towards a deeper understanding
of string-string dualities
\cite{stdua_1,stdua_2,stdua_3,stdua_4,stdua_5,stdua_6,stdua_7} together with 
non--perturbative aspects of field and string theory 
\cite{SW_1,SW_2,SWmore_1,SWmore_2}, has been recently 
achieved within the framework of $N=2$ supersymmetry. Among the reasons for so 
much interest devoted in the last decade to $N=2$ supersymmetric theories is the 
rich geometrical structure of their scalar manifold $\cM_{scalar}$, which has the form:
\begin{equation}
{\cal M}_{scalar} = {\cal SK}_n \, \otimes \, {\cal Q}_m
\label{scalma}
\end{equation}
where ${\cal SK}_n$ denotes a complex $n$--dimensional special K\"ahler
manifold \cite{speckal_1,speckal_2,speckal_3,skgsugra_1,speckal_4} 
(for a review of
Special K\"ahler geometry see either \cite{pietrolectures} or
\cite{toinelectures}) and ${\cal Q}_m$ a $m$--dimensional quaternionic
manifold, $n$ being the number of vector multiplets and $m$ the  number of
hypermultiplets \cite{quatgeom_1,quatgeom_2,quatgeom_3,skgsugra_1}.
\par 
Unfortunately no phenomenological prediction can be drawn at an $N=2$ level,
mainly because of the presence of {\it mirror fermions} which cause these 
theories to be non--chiral. Within the supersymmetry framework, only $N=1$ 
theories meet the phenomenological requirement for a chiral theory. 
In order to extend the results obtained in $N=2$ theories to an $N=1$ level,
much interest has been recently devoted towards finding a mechanism of 
spontaneous $N=2 \to N=1$ supersymmetry breaking. On the other hand, 
there are hints that such 
a spontaneous SUSY breaking should occur from a non--perturbative 
analysis of $N=2$ supergravities based on the study of extremal black hole 
solutions \cite{kalloshlast}.
\par
 Results obtained time ago, in the context of superconformal tensor calculus
lead to a ``no--go'' theorem, stating the impossibility for a spontaneous
 $N=2$ to $N=1$ SUSY breaking to occur \cite{lucia_nogo}. Eventually,
with the developments in special K\"ahler geometry \cite{stdua_4}
\cite{stdua_5}\cite{newspec_1}, this theorem was understood 
to be a consequence of unnecessary restrictions 
imposed on the formulation of special K\"ahler geometry
\cite{n2break_1,n2break_2,n2break_3,n2break_4} (namely the condition of the existence of a holomorphic {\it prepotential} function) and
 could be removed within an extended definition of special K\"ahler manifolds.
As a matter of fact, the more general models without a prepotential can be
formally obtained by combining superconformal tensor calculus (SCT) with
appropriate symplectic transformations which bring SCT models into 
non--physically equivalent ones.
\par
The first achievements in finding a mechanism of
partial supersymmetry breaking have been recently obtained both in a 
global $N=2$ supersymmetric minimal model \cite{n2break_2} and in a 
local one \cite{n2break_1}. The former model can be obtained from the latter by means of a suitable  flat limit. In the local minimal model 
\cite{n2break_1}, in which supergravity was coupled with one vector multiplet 
and one hypermultiplet, a symplectic gauge was
 used which is compatible with the T--duality of string theory, and is 
allowed only in the extended formulation of special K\"ahler geometry. 
\par
In a recent publication \cite{ourn2n1} we succeeded in extending the results obtained in 
\cite{n2break_1} to an $N=2$ supergravity theory coupled to an arbitrary
 number $n$ of vector multiplets and $m$ of hypermultiplets,
breaking spontaneusly to an $N=1$ theory with the survival of an unbroken
 arbitrary compact gauge group (for a general review about 
spontaneous supersymmetry breaking see also \cite{porrati}).
 The formalism adopted for describing a generic 
$N=2$ matter--coupled 
theory is consistent with the conventions of \cite{fundpaper}.
\par
A specific example was worked out \cite{ourn2n1}, corresponding to the choice 
${\cal SK}_n \, = \, SU(1,1)/U(1) \, \otimes \, SO(2,n)/SO(2) \times SO(n)$ for the special K\"ahler manifold and ${\cal Q}_m \, = \,
 SO(4,m)/SO(4) \times SO(m)$ for the quaternionic one.
\section{Partial $N=2$ supersymmetry breaking: general features and 
results}
The partial $N=2$ supersymmetry breaking in a supergravity model is a consequence of 
 a {\it super--Higgs} mechanism and requires the following minimal 
ingredients in order to take place:
\begin{itemize} 
\item{the gauging of a gauge group ${\IR^{*}}^2$ by the graviphoton
 $A_{\mu}^{0}$ and a vector field $A_{\mu}^{1}$}
\item{one hypermultiplet (the {\it hidden sector}) charged with respect 
to the ${\IR^{*}}^2$ group}
\end{itemize}
$N=2$ supersymmetry is violated by spontaneously breaking the $O(2)$ symmetry interchanging
 the two gravitinos. One of the latters acquires mass by ``eating'' a fermion 
field belonging to the vector multiplet of $A_{\mu}^{1}$, and 
 consequently becoming the top spin state of an $N=1$ massive spin--$3/2$ multiplet.
The latter, in order to be completed, needs two massive spin--$1$ fields 
which are provided by the graviphoton $A_{\mu}^{0}$ and $A_{\mu}^{1}$.
These two vector fields acquire mass by ``eating'' two
real scalar fields from the hypermultiplet, by means of an ordinary {\it Higgs}
 mechanism. The remaining spin--$1/2$ Majorana field in the vector multiplet 
fits the lowest spin state of the massive spin--$3/2$ $N=1$ multiplet.
Therefore, as a consequence of this mechanism, we are left with a massless
$N=1$ graviton multiplet, two massless {\it chiral} $N=1$ multiplet and a massive spin--$3/2$ $N=1$ multiplet.
\begin{figure}
\epsfxsize=5.truecm
\epsfysize=5.truecm
\centerline{\hbox{\epsffile{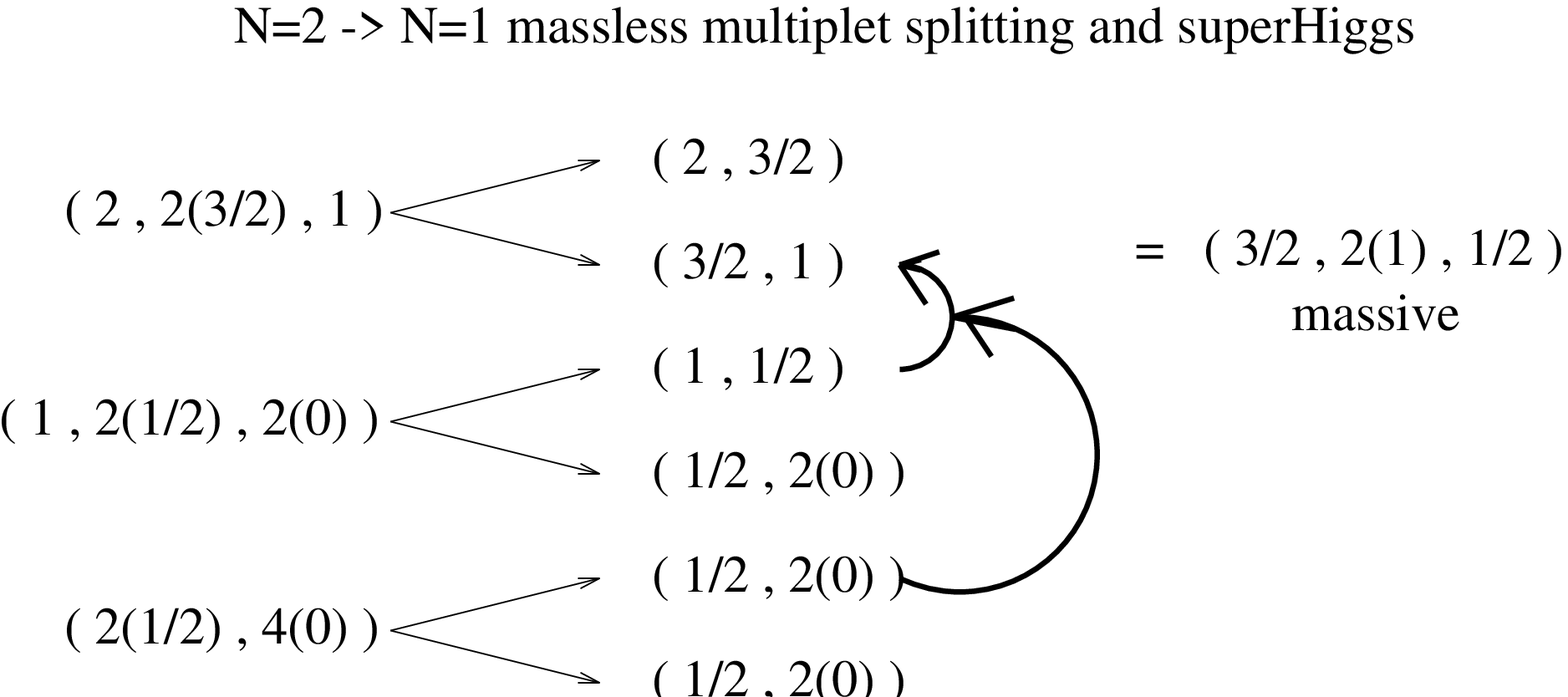}}}
\end{figure}
\par
The mathematical formulation of the problem is based on the following general
 property: in an $N$--extended supergravity theory describing a certain number of scalar fields $\Phi^{I}$, 
 a {\it bosonic background} ($\Phi^{I}_{o}$) admitting $r$ 
{\it killing spinors}
 is not only an extremum (vacuum) of the scalar 
potential but it also breaks $N-r$ of the initial $N$ supersymmetries.
By {\it bosonic background} we mean a state defined by a Minkowskian metric, 
vanishing expectation value of the vector fields and a constant scalar field 
configuration $\Phi^{I}=\Phi^{I}_{o}$. A {\it killing spinor} 
is a constant spinorial parameter such that the corresponding 
supersymmetry transformation of the spinor fields vanishes. Since the
supersymmetry variation of the gravitino on a bosonic background
is proportional to the contraction of 
the gravitino mass--matrix with the SUSY parameter, it is necessary
for the gravitino mass--matrix to have a null eigenvalue 
 in order for a killing spinor to exist.
\par
The occurrence of a partial $N=2\to N=1$ supersymmetry 
breaking depends on the choice of the symplectic gauge for the vector fields,
since different symplectic gauges may lead, after gauging, 
to inequivalent theories \cite{pietrolectures}\cite{fundpaper}. 
\par
As previously pointed out, we solved explicitly the case of an $N=2$ 
supergravity in which the scalars span the manifolds:
${\cal SK}_n \, = \, SU(1,1)/U(1) \, \otimes \, SO(2,n)/SO(2) \times SO(n)$; ${\cal Q}_m \, = \,
 SO(4,m)/SO(4) \times SO(m)$ \cite{ourn2n1}.\\
Within the coordinate--free definition of special K\"ahler manifold \cite{speckal_3}, \cite{skgsugra_1}, we chose the symplectic gauge
corresponding to the Calabi-Visentini coordinates ($S,y^{\alpha}$) 
\cite{pietrolectures}\cite{newspec_1} for 
${\cal SK}_n$, which is compatible with the T--duality symmetry in string theory.
 Moreover we gauged a group of the form:
\begin{equation}
G_{gauge}={\IR^{*}}^{2}\otimes G_{compact}
\end{equation}
where the non--compact factor ${\IR^{*}}^{2}$, responsible for the partial supersymmetry breaking in the way described above, is a subgroup of the isometry group
$\cQ_m$, while the compact factor which will survive as the gauge group at the $N=1$ level, consists of isometries of eighter ${\cal SK}_n$ or $\cQ_m$. 
\par
Denoting by $q^a,\quad (a=1,\cdots 4m)$ the coordinates on $\cQ_m$, for a 
bosonic background defined by a generic point ($S, q^a$) on the $y^{\alpha}=0$
hypersurface of $\cM_{scalar}$, suitable generators of $G_{gauge}$ have been
 found within the isometry algebra of $\cM_{scalars}$, such that a killing 
spinor exists. The generators $T_0,\, T_1$ of the ${\IR^{*}}^2$ factor, which serve the purpose, have, in the {\it canonical} basis of {\bf so(4,m)}, the form:
\begin{eqnarray}
T_{\Lambda}(q)&=&\IL(q) T_{\Lambda}\IL(q)^{-1}\quad \Lambda=0,1\\
T_{0}&=&E_{\epsilon_{l-3}-\epsilon_{l-1}}+
E_{\epsilon_{l-3}+\epsilon_{l-1}}\\
 T_{1}&=&-(E_{\epsilon_{l-3}+
\epsilon_{l}}+E_{\epsilon_{l-3}-\epsilon_{l}})
\end{eqnarray}
where we have denoted by $\IL(q)$ the coset representative on $\cQ_m$.
For this result to be true, the coupling constants $g$ and $g'$,
associated with the ${\IR^{*}}^{2}$ generators, are required to fulfill the condition
$g=g'$, which ensures the existence of a vanishing eigenvalue for the gravitino
mass matrix, the corresponding eigenvector being the killing spinor. The mass 
of the massive gravitino is found to be proportional to $g+g'=2g$.
Therefore the choice of the Calabi--Visentini parametrization 
for the special K\"ahler manifold and a suitable choice of gauging
allowed to define a theory in which partial SUSY breaking $N=2\to N=1$ 
occurs. Moreover this mechanism leaves the gauge symmetry 
corresponding to a $G_{compact}=SO(n-1)$ unbroken at the $N=1 $ level.
\par
A deeper understanding of the physical meaning underlying the 
choice made for the generators of $G_{gauge}$, may be achieved using 
Alekseevskii's representation of quaternionic manifolds
\cite{aleks} \cite{beria}(see also \cite{cecale,cecfergir,vanderseypen}). This 
mathematical formulation allows to describe $\cQ_{m}$ as a group manifold
generated by a {\it solvable} Lie algebra $V_m$ so that the hyperscalars 
are just coordinates parametrizing it. The generators chosen for the group
${\IR^{*}}^{2}$  belong to a {\it maximal abelian ideal} $\cA\subset V_m$ and are
 parametrized by two hyperscalars ($t^0,t^1$) on which the non--compact abelian group act as traslations. It turns out that these two fields 
define flat directions of the 
scalar potential. 
\par
In the light of the results obtained in a recent paper by some of us 
\cite{ramondram},
the nilpotent elements of the maximal abelian algebra $\cA$ are naturally
related to the Peccei-Quinn symmetry generators and 
the two scalars $t^0,t^1$ may be interpreted as Ramond--Ramond fields.
The partial SUSY breaking $N=2\to N=1$ may thus be thought of as an 
effect of the ``condensation'' of the two R--R fields $t^0,t^1$ in the vacuum
state \cite{polstr}\cite{michelson}.
\par
In conclusion, from our analysis it follows that $N=2$ supergravity
can be spontaneously broken to $N=1$ supergravity, with the survival
of unbroken rather arbitrary gauge group. In order to open new possibilities
for phenomenological model building we still have to face the problem 
of breaking the symmetry between the fermions and their {\it mirror} 
fields. It is also an interesting open question to find the relation of our mechanism with the non--perturbative $N=2$
supergravities predicted by string--string
duality and with the conjectured non perturbative breaking caused by extremal black-holes \cite{kalloshlast}.


\begin{thebibliography}{50}
\bibitem{stdua_1}S.~Kachru and C.~Vafa, Nucl. Phys. B450 (1995) 69,
hep-th/9505105.
\bibitem{stdua_2}S.\ Ferrara, J.\ A.\ Harvey, A.\ Strominger and C.\ Vafa,
Phys. Lett. B361 (1995) 59, hep-th/9505162.
\bibitem{stdua_3} M. Bill\'o, R. D'Auria, S. Ferrara, P. Fr\`e, P. Soriani
and A. Van Proeyen, ``R-symmetry and the topological twist of N=2
effective
supergravities of heterotic strings'', preprint hep-th/9505123, to be
published in Int. J. Mod. Phys. {\bf A}.
\bibitem{stdua_4} M. Bill\'o, A. Ceresole, R. D'Auria, S. Ferrara, P. Fr\`e,
 T. Regge, P. Soriani
and A. Van Proeyen, ``A search for non--perturbative dualities
of local N=2 Yang--Mills theories from Calabi--Yau three--folds'',
preprint hep-th 9506075,
to appear on Class and Quant. Grav.
\bibitem{stdua_5}
I.~Antoniadis, S.~Ferrara, E.~Gava, K.S. Narain and T.R. Taylor,
Nucl. Phys. B447 (1995) 35,
hep-th  9504034 (1995).
\bibitem{stdua_6} A. Sen "Strong--Weak coupling duality in
Four--dimensional String Theory" hep-th / 9402002.
\bibitem{stdua_7} A. Sen, Nucl. Phys B388 (1992) 457 and Phys. Lett.
 B303 (1993)
22; A Sen and J.H. Schwarz, Nucl. Phys. B411 (1994) 35; Phys. Lett. B312
(1993) 105.
\bibitem{SW_1}
N.~Seiberg and E.~Witten, Nucl. Phys. B426 (1994) 19.
\bibitem{SW_2}
N.~Seiberg and E.~Witten, Nucl. Phys. B431 (1994) 484.
\bibitem{SWmore_1} A. Klemm, W. Lerche and P. Mayr, Phys. Lett. B357 (1995) 313,
hep-th/9506112.
\bibitem{SWmore_2}A.~Klemm, W.~Lerche and S.~Theisen, ``Nonperturbative Effective
Actions of N=2 Supersymmetric Gauge Theories'', preprint CERN-TH/95-104,
LMU-TPW 95-7, hep-th 9505150.
\bibitem{speckal_1}B. de Wit and A. Van Proeyen, Nucl. Phys. {\bf B245}
(1984) 89;
\bibitem{speckal_2}
B. de Wit, P. G. Lauwers and A. Van Proeyen Nucl. Phys. {\bf B255}
(1985) 569.
\bibitem{speckal_3}
L.~Castellani, R.~D'Auria and S.~Ferrara,
Phys. Lett. 241B (1990) 57,
\bibitem{skgsugra_1}
R.~D'Auria, S.~Ferrara and P.~Fr\'e,
Nucl. Phys. B359 (1991) 705.
\bibitem{speckal_4}
A.~Strominger, Comm. Math. Phys. 133 (1990) 163.
\bibitem{pietrolectures} P.~Fr\'e, Nucl. Phys. B (Proc. Suppl.) 45B,C
\bibitem{toinelectures} A. Van Proeyen, KUL-TF-95-39, Dec 1995. 40pp. Lectures given at ICTP Summer School in High Energy Physics and Cosmology, Trieste, Italy, 12 Jun - 28 Jul 1995, hep-th/9512139
\bibitem{quatgeom_1} J. Bagger, E. Witten, Nucl. Phys. {\bf B 211} (1983)
302.
\bibitem{quatgeom_2} B. de Wit, R. Philippe, S.Q. Su and
A. Van Proeyen Phys. Lett. {\bf B 134} (1984) 37.
\bibitem{quatgeom_3}
K.~Galicki, Comm. Math. Phys. 108 (1987) 117.
\bibitem{lucia_nogo} S. Cecotti, L. Girardello and M. Porrati, Phys. Lett. 145B (1984) 61.
\bibitem{newspec_1}
A.~Ceresole, R.~D'Auria, S.~Ferrara and A.~Van Proeyen,
 Nucl. Phys.  B444 (1995) 92 \\
``On Electromagnetic Duality in Locally
Supersymmetric N=2 Yang--Mills Theory'', preprint CERN-TH.08/94,
hep-th/9412200, Proceedings of the Workshop on Physics from the Planck
Scale to Electromagnetic Scale, Warsaw 1994.
\bibitem{n2break_1} S.~Ferrara, L.~Girardello and M.~Porrati, Minimal Higgs Branch for the breaking of half of the Supersymmetries in N=2 Supergravity, Phys.Lett. B366 :
p.155-159, 1996
\bibitem{n2break_2} I.~Antoniadis, H.~Partouche and T.R.~Taylor,
Spontaneous breaking of N=2 global supersymmetry, Phys.Lett. B372 :p.83-87, 1996
\bibitem{n2break_3} S.~Ferrara, L.~Girardello and M.~Porrati, Spontaneous Breaking
of N=2 to N=1 in Rigid and Local Supersymmetric Theories. hep-th/9512180, 
Phys. Lett. {\bf B376} (1996) 275
\bibitem{porrati}
M.~Porrati, Spontaneous breaking of extended supersymmetry in global and local 
theories, to be published in the proceedings of Spring School and Work-Shop
on String Theory, Gauge Theory and Quantum Gravity, Trieste, Italy, 18-29 March
1996, hep-th/9609073
\bibitem{n2break_4} L.~Alvarez Gaum\'e, J.~Distler and C.~Kounnas, Softly broken N=2 QCD, hep-th/9604004
\bibitem{ourn2n1} P.~Fr\'e, L.~Girardello, I.~Pesando and M.~Trigiante,
Spontaneous $N=2 \to N=1$ Local Supersymmetry Breaking with Surviving Compact
Gauge Groups, hep-th/9607032
\bibitem{fundpaper} L.~Andrianopoli, M.~Bertolini, A.~Ceresole, R.~D'Auria,
S.~Ferrara, P.~Fr\'e and T. Magri,  hep-th/9605032, hep-th/9603004
\bibitem{aleks} D.V. Alekseevskii, Izv. Akad. Nauk. SSSR, Ser. Mat. Tom. 39 (1975) 297
\bibitem{cecale} S. Cecotti, Comm. Math. Phys. 124 (1989) 23
\bibitem{cecfergir} S. Cecotti, S. Ferrara and L. Girardello, IJMP {\bf A4} (1989) 2475
\bibitem{vanderseypen} B. de Wit, F. Vanderseypen and A. Van Proeyen, Nucl. Phys. B400 (1993) 463
\bibitem{kilspinold}
 F. Cecotti, L. Girardello, M. Porrati:  Proc. IX John Hopkins
 Workshop, World Scientific 1985, Nucl. Phys. B 268 (1986) 265,\\
 S. Ferrara and L. Maiani, Proc. V Silarg Symposium, World Scientific
 1986, \\
For a general discussion see for instance the book\\
L.~Castellani, R.~D'Auria, P.~Fr\'e:
``Supergravity and Superstring Theory: a geometric perspective''. World Scientific 1990, page 1078
and following ones (in particular eq. (IV.8.1)).
\bibitem{beria} V.A. Tsokur and  Yu.M.Zinoviev, N=2 Supergravity models based on the nonsymmetric quaternionic manifolds.1.Symmetries and Lagrangian, hep-th/9605134,\\ V.A. Tsokur and  Yu.M.Zinoviev, N=2 Supergravity models based on the nonsymmetric quaternionic manifolds.2.Gauge interactions, hep-th/9605143  
\bibitem{kalloshlast}
R. Kallosh, Superpotential from black holes, hep-th/9606093
\bibitem{polstr}
J.~Polchinski and A.~Strominger, New Vacua for Type II String Theory,
hep-th/9510227
\bibitem{michelson}
J.~Michelson, Compactifications of type IIB strings to four--dimensions
with non trivial classical potential, hep-th/9610151
\bibitem{ramondram}
L.~Andrianopoli, R.~D'Auria, S.~Ferrara, P.~Fre' and M.~Trigiante,
``R--R scalars, U--duality and Solvable Lie Algebras'', hep-th/9611014
\end{thebibliography}
\end{document}